\documentclass[aps, pra, showpacs, reprint, preprintnumbers,amsmath,amssymb,superscriptaddress,floatfix]{revtex4-2}
\usepackage{graphicx}      
\usepackage{bm}             	
\usepackage{times}			
\usepackage{tabularx}
\usepackage{color}
\usepackage{dcolumn}		
\usepackage{amsmath}
\usepackage{amssymb}
\usepackage{amsfonts}
\usepackage{times}
\usepackage{xcolor,colortbl}
\usepackage{amsmath}
\usepackage{multirow}
 \usepackage{ragged2e}
\usepackage[T1]{fontenc}
\usepackage[autostyle]{csquotes}
   \MakeAutoQuote{‘}{’}
\usepackage{csquotes} 
\usepackage[english]{babel}
\definecolor{lime}{HTML}{A6CE39}
\usepackage{orcidlink}
\usepackage[normalem]{ulem}
\let\cite\onlinecite

\makeatletter
\newcommand\xleftrightarrow[2][]{%
  \ext@arrow 9999{\longleftrightarrowfill@}{#1}{#2}}
\newcommand\longleftrightarrowfill@{%
  \arrowfill@\leftarrow\relbar\rightarrow}
\makeatother
\setcitestyle{square}

\begin{document}
\title{Micromagnetic Modeling of Surface Acoustic Wave Driven Dynamics: Interplay of Strain, Magnetorotation, and Magnetic Anisotropy}
\author{Florian Millo \orcidlink{0000-0002-8248-6635}}
\email{florian.millo@insp.jussieu.fr}
\affiliation{Sorbonne Universit\'e, CNRS, Institut des Nanosciences de Paris, F-75005 Paris, France}
\author{Pauline Rovillain \orcidlink{0000-0002-3960-4544}}
\affiliation{Sorbonne Universit\'e, CNRS, Institut des Nanosciences de Paris, F-75005 Paris, France}
\author{Massimiliano Marangolo \orcidlink{0000-0001-6211-8168}}
\affiliation{Sorbonne Universit\'e, CNRS, Institut des Nanosciences de Paris, F-75005 Paris, France}
\author{Daniel Stoeffler \orcidlink{0000-0002-1637-0183}}
\affiliation{Universit\'e de Strasbourg, CNRS, Institut de Physique et Chimie des Mat\'eriaux de Strasbourg, UMR 7504, France}
\date{\today}                                           
\begin{abstract}
We study the coupling mechanism of surface acoustic waves (SAW) with spin waves (SW) using micromagnetic analysis. The SAW magnetoacoustic excitation field is fully implemented, i.e., all strain and lattice-rotation terms are included. A realistic CoFeB film with a weak in-plane uniaxial anisotropy is considered. We investigate the conditions for efficient SAW--SW coupling, with particular emphasis on the case where the SAW propagates parallel to the external magnetic field, a configuration of special interest for magnonic applications. Remarkably, we find that the anisotropy orientation serves as a knob to tune the parallel resonant interaction. Overall, this work provides a unified and practical picture of SAW--SW coupling in thin magnetized films.
\end{abstract}

\maketitle
\begin{figure}[!t]
    \centering
    \includegraphics[width=0.79\linewidth]{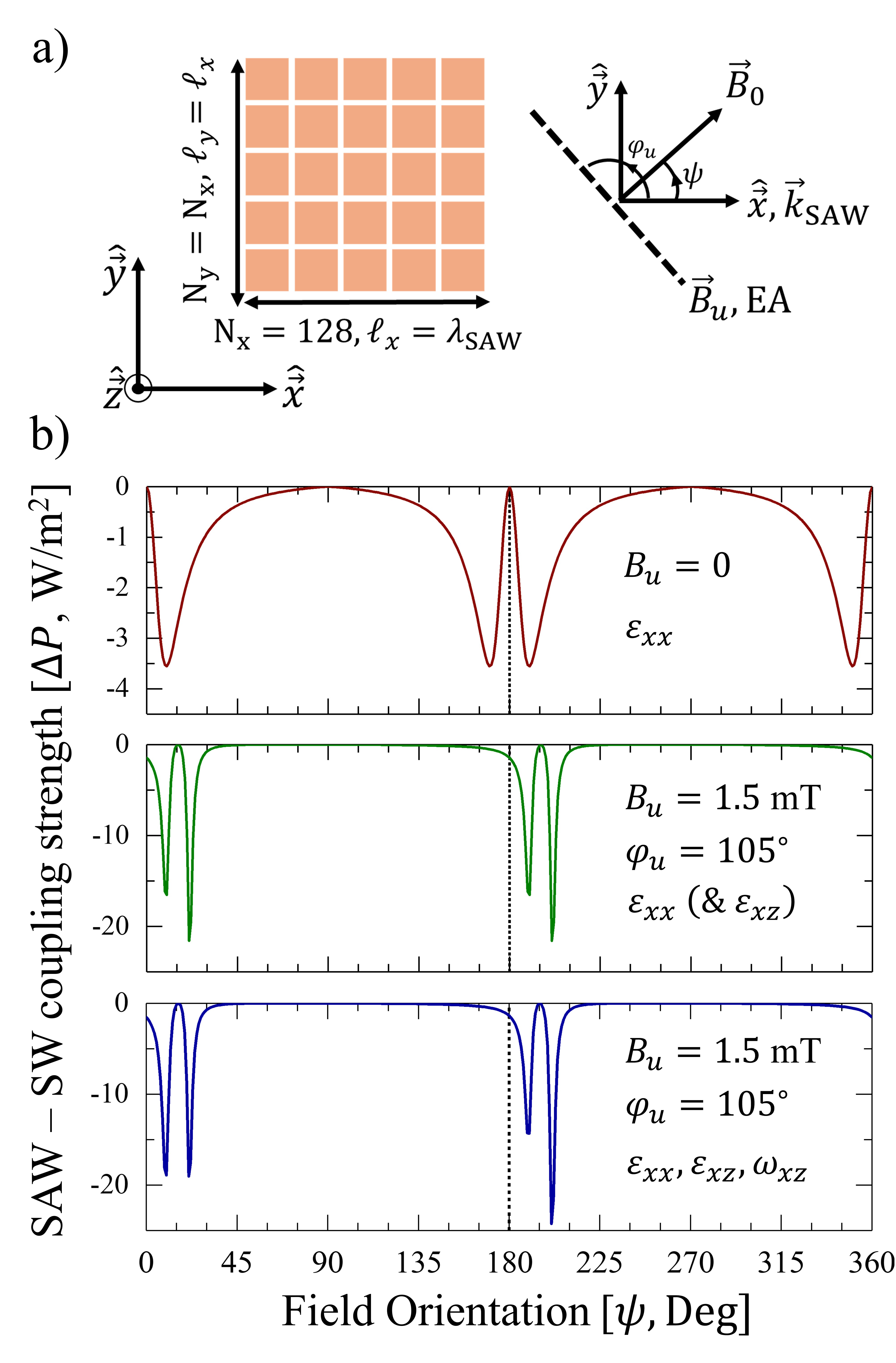}
    \caption{a) Geometry of the problem and definitions. The slab has dimensions of $\ell_x=\ell_y= \lambda_{\textrm{SAW}}$, $\ell_z=34$ nm and is meshed into $\{N_x,N_y,N_z\}=\{128,128,1\}$. An in-plane static field $\vec{B}_0$ is applied with an angle $\psi$ with respect to $\vec{k}_{\textrm{SAW}} = k_\textrm{{SAW}}.\hat{\vec{x}}$. A weak uniaxial anisotropy field $\vec{B}_{u}$ with an angle $\varphi_u$ is introduced in the system. EA stands for easy axis. b) Symmetry of the SAW--SW coupling strength $\Delta P(\psi)$ at $f_{\textrm{SAW}}=1.72$ GHz and $B_0=1.5$ mT. The material parameters used in the simulations are as follows. SAW parameters: velocity $v_{\textrm{SAW}}=3870$ m/s, wavevector $|\vec{k}_{\textrm{SAW}}| = 2.8$ rad/$\mu$m, longitudinal strain $\varepsilon_{xx}=0.75 \times10^{-5}$, shear strain $\varepsilon_{xz}=0.05\times10^{-5}$, lattice rotation $\omega_{xz}=1\times10^{-5}$ [\cite{lopes_seeger_symmetry_2024}]; Magnetics: Saturation magnetization $M_s=1.35$ MA/m, Exchange coupling $A_{\textrm{ex}}= 21$ pJ/m, Gilbert damping $\alpha_G=0.01$, Shape anisotropy $B_{\textrm{shape}}=1$ mT, Magnetoelastic coupling $B_1= B_2=-7.6 $ MJ/m$^3$. The vertical dashed line marks $\psi=180^\circ$.}
    \label{fig1:method}
\end{figure}

 
The coupling between surface acoustic waves (SAWs), i.e., mechanical strains, and magnetic eigen-excitations [\cite{kittel_excitation_1958, akhiezer_coupled_1958, tiersten_coupled_1964, bommel_excitation_1959, kobayashi_ferromagnetoelastic_1973}] has attracted increasing attention in recent years in the field of spintronics [\cite{Flebus_2024, Krenner_2026}]. A particular type of SAW is the Rayleigh SAW [\cite{rayleigh_waves_1885}], a propagating mechanical strain with a well-defined frequency ($f_{\textrm{SAW}}$) and wavevector ($\vec{k}_{\textrm{SAW}}$). Due to symmetry relations, the Rayleigh SAW couples to propagating magnetic eigen-excitations i.e., spin waves (SW)  [\cite{tiersten_coupled_1964, bommel_excitation_1959,kobayashi_ferromagnetoelastic_1973, weiler_elastically_2011}]. Hence, the Rayleigh SAW can act as a natural moving drive for SWs via an effective magnetoacoustic excitation field $\vec{B}_\textrm{{exc}}$ reflecting both contributions of the strain tensor $\varepsilon_{\mu \nu} = (\partial_\nu u_\mu + \partial_\mu u_\nu)/2$ and lattice rotation tensor $\omega_{\mu\nu}= (\partial_\nu u_\mu - \partial_\mu u_\nu)/2$ where $u$ is the SAW displacement vector and $\{\mu,\nu=x,y,z\}$. SWs in in-plane magnetized films are intrinsically anisotropic (dipolar-dominated [\cite{kalinikos_theory_1986}]) displaying a twofold symmetry in their angular response. 
The introduction of a weak in-plane uniaxial anisotropy produces a competing symmetry axis that reorients the magnetization equilibrium and, in turn, reshapes the SW spectrum  such that the magnetoacoustic effective field couples differently to the SW eigenmodes.\\
\indent A large fraction of prior works have been formulated within formalisms [\cite{weiler_elastically_2011, dreher_surface_2012, thevenard_surface-acoustic-wave-driven_2014,  
labanowski_power_2016, 
sasaki_nonreciprocal_2017, xu_nonreciprocal_2020, hernandez-minguez_large_2020, Puebla_2020, babu_interaction_2021,GaoRunliang2022,
chen_widen-dynamic-range_2024, lopes_seeger_symmetry_2024, millo_symmetry_2025}], which are powerful for describing uniform responses and static dipolar interaction [\cite{gowtham_traveling_2015, rovillain_impact_2022, yamamoto2022, Hu2023, yamamoto_magnetostatic_2023, huang_large_2024, Vythelingum2025, Sharma2026}] but cannot resolve \textit{dynamic} $\vec{k}$-dependent dipolar interaction  --  requiring an exact treatment of the dipolar energy rather than an approximation [\cite{kalinikos_theory_1986, vanderveken_confined_2021}] -- or the spatial structure of propagating SW eigenmodes. In recent works, a micromagnetic analysis of SAW-driven parallel spin pumping of forward volume mode [\cite{Jander2025}], as well as practical implementations of strain driven terms in MuMax3 [\cite{vansteenkiste_design_2014, moreels2025mumaxextensiblegpuacceleratedmicromagnetics}], which both enhance and facilitate the processing of magnetoelastic coupling for various experimental configurations have been proposed [\cite{yemeli2025}]. Extending these micromagnetic investigations is necessary to properly quantify the SAW--SW interaction in various experimental geometries and to design magnonic devices controlled by SAWs rather than by inductive antennas [\cite{yemeli2025}].\newpage
\indent In this Letter, we investigate through a micromagnetic analysis the SAW--SW coupling dynamics by implementing a magnetoacoustic excitation field where all strain and lattice rotation terms are included. Importantly, a realistic CoFeB film with a weak in-plane uniaxial anisotropy is taken into consideration. This work aims to determine the conditions for efficient SAW--SW coupling by taking into account the intensity of the external field, all SAW contributions (strain and magnetorotation) and/or the sample's magnetic anisotropy. In particular, we show that the orientation of the anisotropy can be used as a knob to enhance the parallel SAW--SW coupling i.e., when the applied dc field is collinear with the SAW wavevector. This geometry is often considered for magnonic and sensing applications [\cite{Krenner_2026, rovillain_impact_2022}]. 
Building on a micromagnetic analysis where all $\{\varepsilon_{\mu\nu},\omega_{\mu\nu}\}$ terms are incorporated, and using the uniaxial anisotropy as a knob to reshape the parallel SAW--SW interaction, this analysis provides a unified and practical picture for SAW--SW coupling dynamics in thin magnetized films.\\
\indent We perform a micromagnetic analysis -- based on the GPU-accelerated MuMax3 software [\onlinecite{vansteenkiste_design_2014}] -- to calculate the SAW--SW coupling dynamics. To mimic experiments performed on CoFeB/LiNbO$_3$ [\onlinecite{millo_symmetry_2025}], for $|\vec{k}_{\textrm{SAW}}| \neq 0$, we first define a slab of in-plane dimensions $\{\ell_x \times \ell_y\}$ with $\ell_x=\ell_y=\lambda_{\textrm{SAW}}$ (where $\lambda_{\textrm{SAW}}$ is the SAW wavelength) and thickness $\ell_z=34\;$nm. The slab is replicated using periodic boundary conditions (PBC) with $\textrm{PBC}_x = \textrm{PBC}_y = 100$ to simulate an infinite film. We reduced the computation time by discretizing the slab into $128\times128\times1$ cells [\footnote{We verified that discretizing the thickness with $N_z = 8$ cells does not improve the result as compared to $N_z = 1$.}]. An in-plane static field $\vec{B}_0$, in increments of $0.1$ mT, is applied with an angle $\psi$ with respect to $\vec{k}_{\textrm{SAW}} = k_\textrm{{SAW}}.\hat{\vec{x}}$ and angular studies of the coupling dynamics are performed by varying the angle $\psi$ in increments of $1^\circ$ [Fig.~\ref{fig1:method}(a)]. For each pair of parameters $\{\vec{B}_0,\psi\}$, the system is first relaxed to its equilibrium state. The magnetoacoustic excitation is applied as a propagating wave for both $\varepsilon_{\mu\nu}$ and $\omega_{\mu\nu}$ terms,
\begin{align}
   \left\{ \begin{matrix} &\varepsilon_{\mu\nu}(\vec{r},t) = \varepsilon_{\mu\nu}^0\cos(\vec{k}_{\textrm{SAW}}\cdot\vec{r} - 2 \pi f_{\textrm{SAW}} t-\Delta_{\mu\nu}) \label{eps_omega},\\
    &\omega_{\mu\nu}(\vec{r},t) = \omega_{\mu\nu}^0\sin(\vec{k}_{\textrm{SAW}} \cdot \vec{r} - 2 \pi f_{\textrm{SAW}} t+\Delta_{\mu\nu}). \end{matrix} \right.
\end{align}\newpage
In Eq.~\ref{eps_omega}, the $\pi/2$ dephasing between strain $\varepsilon_{\mu \nu}$ and lattice rotation $\omega_{\mu\nu}$ is taken into consideration. Moreover, the dephasing of the diagonal and off-diagonal terms of strain and lattice rotation are taken into account through the $\Delta_{\mu\nu}=0\;(\pi/2)$ for $\mu=\nu\; (\mu\neq\nu)$, respectively. The values of $\varepsilon_{\mu \nu}^0$ and $\omega_{\mu\nu}^0$ are taken from ref.~[\cite{lopes_seeger_symmetry_2024}]. 
The magnetoacoustic excitation field $\vec{B}_\textrm{{exc}}(\vec{r},t)$ is then equal to the sum of the usual magnetoelastic and the magnetorotation terms,
\begin{equation}
\resizebox{0.91\linewidth}{!}{$
\begin{aligned}
    \vec{B}_\textrm{{exc}}(\vec{r},t) &= -\frac{2}{M_s} \left( B_1  \begin{bmatrix} \varepsilon_{xx}m_x \\ \varepsilon_{yy}m_y \\ \varepsilon_{zz}m_z \end{bmatrix} +B_2 \begin{bmatrix} \varepsilon_{xy}m_y + \varepsilon_{xz}m_z\\ \varepsilon_{xy}m_x + \varepsilon_{yz}m_z \\ \varepsilon_{xz}m_x + \varepsilon_{yz}m_y \end{bmatrix} \right) \\
    &- \mu_0 M_s \begin{bmatrix} \omega_{xz} m_z \\ \omega_{yz} m_z \\\omega_{yz} m_y  +\omega_{xz} m_x  \end{bmatrix},
\end{aligned}
$}
\label{Bexc_mel}
\end{equation}
where $B_1$ and $B_2$ are the magnetoelastic coupling constants. In Eq.~\ref{Bexc_mel}, we have neglected the anisotropy terms of the magnetorotational coupling [\cite{xu_nonreciprocal_2020, yamamoto2022}]. In addition, we assume that the CoFeB material is an isotropic magnetic system, hence $B_1 = B_2$. The SAW attenuation into the magnetic film is quantified by the magnetic power absorption [W/m$^2$], 
\begin{align}
    \Delta P(f) = -\pi f \ell_z \Im\{\tilde{\vec{B}}^{\dagger}_{\textrm{exc}}(f) \cdot \overleftrightarrow \chi(f) \cdot  \tilde{\vec{B}}_{\textrm{exc}}(f)\}\label{magnetipowerabsorption1}
\end{align}
where tilde quantities are the complex Fourier amplitudes of quantities in the frequency domain, $\chi$ is the Polder tensor and $(.)^{\dagger}$ denotes the conjugate transpose operation. We rewrite Eq.~\ref{magnetipowerabsorption1} in terms of real and imaginary components of the complex amplitudes extracted from the simulations,
\begin{equation}
\resizebox{0.9\linewidth}{!}{$
\begin{aligned}
        \Delta P(f) = - \pi f M_s \ell_z 
        \sum_{\alpha\in {x,y,z}} \left( 
        \tilde{B}_{\textrm{exc},\alpha,r}\delta \tilde{m}_{\alpha,i} - \tilde{B}_{\textrm{exc},\alpha,i}\delta \tilde{m}_{\alpha,r} 
        \right),
\end{aligned}$}
\label{magnetipowerabsorptionfinaldetailled}
\end{equation}
where $\delta \tilde{m}$ is the reduced magnetization in the frequency domain, and the subscripts $r\;\textrm{and}\;i$ denote the real and imaginary parts, respectively. The step-by-step derivation of Eq.~\ref{magnetipowerabsorptionfinaldetailled} from Eq.~\ref{magnetipowerabsorption1} is given in Appendix A. The post-processing, including the Fourier transforms, is performed within the MuMax3 run. The magnetic order parameter exhibits a transient regime at early times, which evolves into a stationary regime within a few nanoseconds (not shown). Once the stationary regime is established, SAWs become the sole excitation source for the magnetization order parameter. We then use the stationary regime to calculate $\Delta P (f)$ [Fig.~\ref{fig1:method}(b)].
\\
\indent The aim of our simulations is to show how the magnetorotational term and the magnetic anisotropy term affect the dynamics of the SAW--SW interaction. In Fig.~\ref{fig1:method}(b), we first compare the case where only the longitudinal strain term is considered in an isotropic medium with a simulation that includes magnetic anisotropy. We then introduce the shear strain term and the magnetorotational term to highlight their additional impact on the coupled dynamics.
To isolate the role of the uniaxial anisotropy, we first consider a configuration where only the longitudinal strain component $\varepsilon_{xx}$ is applied. The resulting angular dependence of the SAW--SW coupling strength $\Delta P(\psi)$ at $f_{\mathrm{SAW}} = 1.72$~GHz [\footnote{The choice of SAW frequency ($f_{\mathrm{SAW}} = 1.72$~GHz) follows from experimental measurements in ref.~[\cite{millo_symmetry_2025}]}] and $B_0 = 1.5$~mT is shown in Fig.~\ref{fig1:method}(b).
In the absence of uniaxial anisotropy ($B_u = 0$ and $\varepsilon_{xx}$, top panel), $\Delta P(\psi)$ displays a smooth and symmetric angular variation with zero coupling (no parallel SAW--SW interaction) when $\vec{B}_{0}\parallel\vec{k}_{\textrm{SAW}}$. Interestingly, the usual fourfold symmetry of $\varepsilon_{xx}$ [\cite{weiler_elastically_2011}] is absent here. This absence arises from the interplay between dipolar symmetries ($ \propto \sin^2(\psi)$, \cite{kalinikos_theory_1986}) and
$\varepsilon_{xx}\propto\sin(2\psi)$. For $f_{\textrm{SAW}}=1.72$ GHz, the SAW--SW coupling occurs when the static magnetization is nearly collinear with the SAW wavevector (see for e.g. angular dependence of in-plane SWs in Fig.~2(b) of ref.~[\cite{millo_symmetry_2025}]). When a weak in-plane uniaxial anisotropy field $B_u = 1.5$~mT is introduced at an orientation $\varphi_u = 105^{\circ}$ [\cite{millo_symmetry_2025}], the line shape of $\Delta P(\psi)$ is modified (interplay of $B_u=1.5$ mT, $\varepsilon_{xx}$ and dipolar terms, middle panel): sharp, narrow and intense minima emerge near the hard axis direction. We find that SAW--SW coupling strength $\Delta P$ is no longer zero when $\vec{B}_0\parallel \vec{k}_{\textrm{SAW}}$ i.e., a parallel SAW--SW interaction is obtained.
\begin{figure*}[!t]
    \centering
    \includegraphics[width=0.85\linewidth]{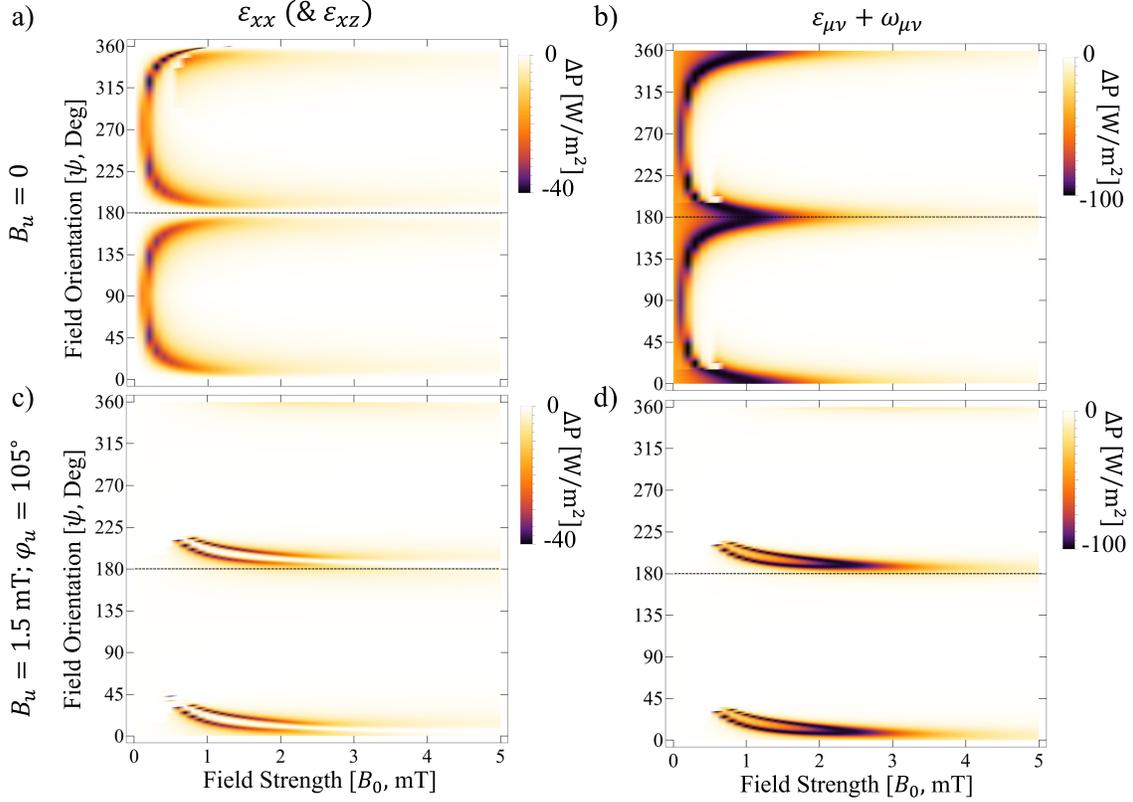}
   \caption{2D maps of the SAW--SW coupling strength $\Delta P(B_0,\psi)$ highlighting the interplay of anisotropy with magnetoacoustic excitation. Left column: $\varepsilon_{xx}$ (\& $\varepsilon_{xz}$). Right column: full magnetoacoustic drive including strain and rotation ($\varepsilon_{\mu\nu}\,\textrm{and}\,\omega_{\mu\nu}$). (a,b) In the absence of anisotropy $B_u=0$. (c,d) Weak anisotropy $B_u=1.5$~mT oriented at $\varphi_u=105^\circ$. The horizontal dashed line marks $\psi=180^\circ$.}
    \label{fig2:SAWSWBu0Buneq0}
\end{figure*}
We then include the shear strain $\varepsilon_{xz}$, but due to its smallness with respect to $\varepsilon_{xx}$, its addition did not change the power absorption map. 
Moreover, one can observe a nonreciprocal SAW propagation, i.e. the symmetry between the parallel and antiparallel configurations of $k_{\textrm{SAW}}$ and the applied dc field is lost.
Remarkably, the addition of lattice rotation $\omega_{xz}$ changed the power absorption map (interplay of $B_u=1.5$ mT, $\varepsilon_{xx}, \varepsilon_{xz}$, $\omega_{xz}$ and dipolar terms, bottom panel) because it introduces an additional magnetization-driving torque beyond the purely magnetoelastic strain channel.\\ \indent As seen in the bottom panel of Fig.~\ref{fig1:method}, $\omega_{xz}$ enhances and reshapes the narrow minima in $\Delta P(\psi)$ near the magnetization reorientation sector (around $\psi\simeq 200^\circ$), producing deeper absorption and additional fine structure compared to the case with only $\varepsilon_{xx}$ (\& $\varepsilon_{xz}$). This indicates that the antisymmetric displacement-gradient component provides a phase-shifted torque that becomes particularly effective when the energy landscape is bistable due to the competition between $B_0$ and the weak uniaxial anisotropy $B_u$. The introduction of $\omega_{xz}$ clearly enhances nonreciprocal features around the magnetization reorientation sector (cf middle and bottom row). This feature has already been studied in refs.~[\cite{xu_nonreciprocal_2020, rovillain_impact_2022}]. \\
\indent The impact of anisotropy, strain, lattice rotations and dipolar terms on the SAW--SW coupling is summarized in Fig.~\ref{fig2:SAWSWBu0Buneq0} through 2D maps of $\Delta P(B_0,\psi)$. For $\{B_u=0,\varepsilon_{xx}$ (\& $\varepsilon_{xz})\}$ [panel (a)], the response exhibits a broad absorption envelope at low fields together with a nearly symmetric angular structure under $\psi\rightarrow \psi+180^\circ$, consistent with a configuration where the equilibrium magnetization follows the applied dc field and the dominant symmetry is set by the relative orientation between $\vec{B}_0$ and $\vec{k}_{\rm SAW}$. When all strain and rotation terms are included [panel (b)], the overall angular symmetry remains, but the coupling strength is strongly enhanced (note the wider color scale reaching $-100$~W/m$^2$) and the absorption region becomes sharper and more structured, particularly around the $\psi=180^\circ$ (horizontal dashed line). This indicates that, even in the absence of uniaxial anisotropy, introducing the lattice rotation component activates an additional driving field that increases the efficiency of magnetoacoustic pumping across a wider set of $\{B_{0},\psi\}$. It is worth emphasizing that the parallel SAW--SW interaction occurs each time the horizontal line is crossed.
\begin{figure*}[!t]
    \centering
    \includegraphics[width=0.85\textwidth]{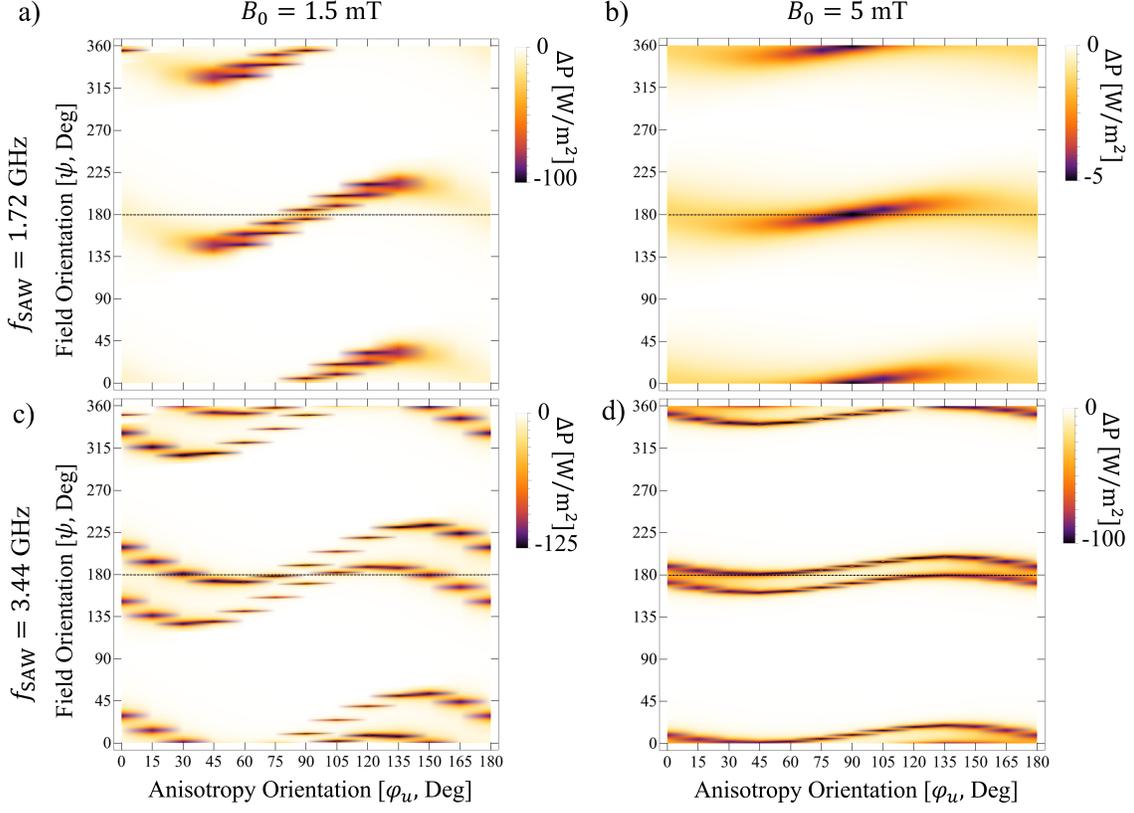}
    \caption{2D maps of the SAW--SW coupling strength $\Delta P(\psi,\varphi_u)$, where $\varphi_u$ is varied in increments of $15^\circ$ and $\psi$ in increments of $1^\circ$, at fixed uniaxial anisotropy field $B_u=1.5$~mT, including full magnetoacoustic drive, with strain and rotation ($\varepsilon_{\mu\nu}\,\textrm{and}\,\omega_{\mu\nu}$). Polynomial nth-order interpolation is applied between simulated $\varphi_u$-curves. Left column: $B_0=1.5$ mT. Right column: $B_0=5$ mT. (a,b) $f_{\textrm{SAW}}=1.72$ GHz. (c,d) $f_{\textrm{SAW}}=3.44$ GHz. The horizontal dashed line marks $\psi=180^\circ$.}
    \label{fig3:SAWSWAnisotropyOrientation}
\end{figure*}
When a uniaxial anisotropy is introduced $\{B_u=1.5$~mT, $\varphi_u=105^\circ\}$ [panel (c) and (d) of Fig.~\ref{fig2:SAWSWBu0Buneq0}], the qualitative behavior of the response changes markedly. In the $\varepsilon_{xx}$ (\& $\varepsilon_{xz}$) configuration [panel (c)], sharp, narrow absorption ridges emerge and form branch-like structures in the $(B_0,\psi)$ plane. These ridges appear only above a finite threshold field (visibly starting around $B_0\approx0.5\;\textrm{mT}$, depending on $\psi$) and follow the angles where the equilibrium magnetization undergoes a rapid reorientation between competing minima set by the Zeeman term and anisotropy. When $B_0$ increases, the uniaxial term becomes progressively less dominant and the reorientation window narrows, leading to a gradual weakening and eventual fading of the SAW--SW coupling strength. Finally, including $\omega_{\mu\nu}$ together with the strain components [panel (d)] amplifies these anisotropy-induced structures: the ridges become deeper, persist over a broader range of $B_0$, and acquire additional fine structure near the resonance sector. This behavior is consistent with the interpretation already suggested in Fig.~\ref{fig1:method}(b): the lattice rotation $\omega_{xz}$ (the most dominant component of $\omega_{\mu\nu}$) provides a phase-shifted torque contribution that becomes particularly effective precisely where the magnetic energy landscape is bistable or soft, i.e., where small changes in $\psi$ or $B_0$ produce large changes in equilibrium magnetization. \\
\indent Overall, Fig.~\ref{fig2:SAWSWBu0Buneq0} shows that the sharp branch features are sensitive to anisotropy, since they are absent for $B_u=0$ and appear abruptly once $B_u\neq 0$. At the same time, the comparison between panels (c) and (d) shows that lattice rotation does not merely rescale $\Delta P$: it selectively enhances and restructures the coupling in the resonance sector, making the SAW--SW interaction more pronounced and more robust in $B_0$.  
Practically, Fig.~\ref{fig2:SAWSWBu0Buneq0} already delivers a key message: the weak uniaxial anisotropy determines where the ridges appear in $\psi$--$B_0$ plane, while lattice rotation determines how strongly they develop (ridge depth i. e., $\forall\psi \textrm{\;what is the} \max[\Delta P(B_0)]$ and persistence $\Delta P(\psi)$ for given $B_0$ range), providing routes to engineer the magnetoacoustic parallel SAW--SW coupling regimes via anisotropy orientation and SAW polarization.\\
\indent In the following, we demonstrate another practical use of our code, which allows us to predict the magnetic anisotropy that should be induced in the sample in order to enhance the SAW--SW coupling in the parallel configuration ($\vec{B}_0\parallel \vec{k}_{\textrm{SAW}}$). We therefore keep the magnitude of the anisotropy field vector constant (i.e. $B_u=1.5$ mT) while varying its orientation, $\varphi_u$. The goal is to enhance the parallel SAW--SW coupling regime, which was still very weak in the bottom panel of Fig.~\ref{fig1:method}(b). The results for two different values of $B_0$ and $f_{\textrm{SAW}}$ are reported in Fig.~\ref{fig3:SAWSWAnisotropyOrientation}.
\begin{figure*}[!t]
    \centering
    \includegraphics[width=0.85\linewidth]{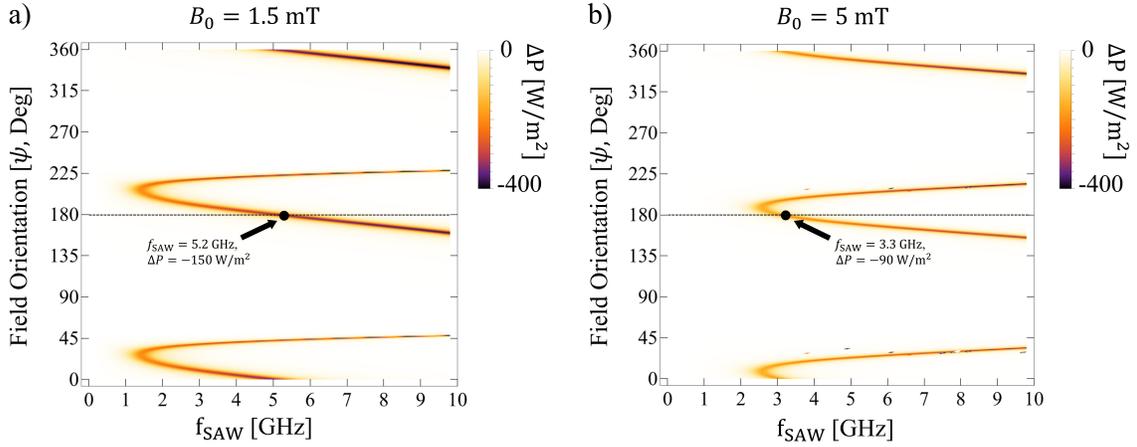}
    \caption{ 2D maps of the SAW--SW coupling strength $\Delta P(f_{\textrm{SAW}},\psi)$ for $B_u=1.5$ mT and $\varphi_u=120^\circ$. a) field strength $B_0=1.5$ mT.  b) $B_0=5$ mT. Parameters used in this simulation are: $|\vec{k}_{\textrm{SAW}}|=2\pi f_{\textrm{SAW}}/v_{\textrm{SAW}}$, full magnetoacoustic drive including strain and rotation ($\varepsilon_{\mu\nu}\,\textrm{and}\,\omega_{\mu\nu}$). The horizontal dashed line marks $\psi=180^\circ$. The arrows show the intersection at which the parallel SAW--SW interaction occurs.}
    \label{fig4:SAWSWParallelPumpingPsivsFsaw}
\end{figure*}
For $\{B_0=1.5\;\textrm{mT}, f_{\textrm{SAW}}=1.72\;\textrm{GHz}\}$ [panel (a)], the parallel SAW--SW interaction vanishes. In contrast, for larger dc fields $B_0=5$ mT [panel (b)], the SAW--SW interaction emerges at $\psi=n\times180^\circ$ with $\Delta P\sim-5$ W/m$^2$. Increasing the SAW frequency to $f_{\textrm{SAW}}=3.44$ GHz [bottom row], yields the following behavior: For $B_{0}=1.5$ mT [panel (c)] the parallel SAW--SW interaction is observed at $\varphi_u=\{30^\circ, 75^\circ, 105^\circ, 150^\circ\}$ for $\psi=\{0^\circ,180^\circ\}$. Rotating $\varphi_u$ changes the effective longitudinal versus transverse components of the magnetoacoustic excitation field.
This $\varphi_u$ rotation shifts both $\Delta P = 0$ and $\Delta P \neq 0$ conturs, thereby reshaping the phase-matching conditions. For $B_{0}=5 \,\textrm{mT} > B_{u}$ [panel (d)], the intense parallel SAW--SW interaction is observed at $\varphi_u=\{30^\circ,45^\circ, 135^\circ,150^\circ\}$ for $\psi=\{0^\circ,180^\circ\}$. Moreover, $\Delta P$ recovers its expected twofold symmetry [\cite{kalinikos_theory_1986, millo_symmetry_2025}]. Put simply, at high SAW frequencies, the SAW--SW coupling $\Delta P$ strongly depends on the geometry of the magnetoacoustic excitation field and the orientation of the anisotropy. \\
\indent Expanding on the results given in Figs.~\ref{fig2:SAWSWBu0Buneq0}, \ref{fig3:SAWSWAnisotropyOrientation}, our micromagnetic analysis provides several actionable consequences for fundamental spintronics applications. Antenna-less SW excitation can be realized via magnetoacoustic drive as explained in ref.~[\cite{yemeli2025}]. Fundamentally, operating in systems with large ellipticity (i.e., dipolar--dominated) minimizes the required SAW--SW coupling strength. Rotating $\varphi_u$ continuously shifts the $\Delta P$ resonance sectors. The leverage of this process is strongest when the field strength is comparable to the uniaxial anisotropy field i.e., $B_0\!\sim\!B_u$. Increasing the SAW frequency (higher $k_{\mathrm{SAW}}$) broadens the $\Delta P$ branch-cut bands and can strengthen coupling [Fig.~\ref{fig3:SAWSWAnisotropyOrientation}(a,c)], so higher SAW harmonics enlarge the SAW--SW interaction window. \\
\indent In Fig.~\ref{fig4:SAWSWParallelPumpingPsivsFsaw}, the phase-matching conditions where the SAW dispersion intersects accessible SW branches at finite $\vec{k}_{\mathrm{SAW}}$ are provided for $B_u=1.5\;\textrm{mT}$ and $\psi=120^\circ$. This indicates a direct ``blueprint'' of the $\{\psi,f_{\mathrm{SAW}}\}$ combinations that maximize power transfer. At $B_0=1.5$~mT [Fig.~\ref{fig4:SAWSWParallelPumpingPsivsFsaw}(a)], $\Delta P$ exhibits a characteristic ``cusp''-like onset about $f_{\mathrm{SAW}}\sim 1$--$2$~GHz from which two prominent ridges emerge and split as $f_{\mathrm{SAW}}$ varies. One ridge family is centered around the so-called backward-volume configuration ($\psi\simeq 0^\circ$ and its periodic counterpart near $\psi\simeq 360^\circ$), while a second family appears around the opposite propagation direction (near $\psi\simeq 180^\circ$). The fact that each ``cusp'' opens into two branches reflects that, close to the field regime $B_0\sim B_u$, small changes in $\psi$ substantially reorient magnetization and hence modify the effective SW dispersion and the phase-matching condition. In other words, the anisotropy rescales $\Delta P$ and reshapes the efficient SAW--SW coupling. When $B_0=5$~mT [Fig.~\ref{fig4:SAWSWParallelPumpingPsivsFsaw}(b)], the ridge pattern becomes smoother and more collimated in $\psi$, consistent with a regime where the Zeeman energy always dominates ($B_0\gg B_u$) and the equilibrium magnetization follows the external field more closely. Correspondingly, the cusp-like onsets shift to slightly higher frequencies and the ridges broaden into linear bands, indicating that the phase-matching is now primarily controlled by the SW dispersion rather than by near-critical reorientation effects. The $180^\circ$ periodicity in $\psi$ remains apparent if we neglect non-reciprocity, as expected for reversing the relative orientation between $\vec{k}_{\mathrm{SAW}}$ and $\vec{B}_0$, and thus provides a robust experimental fingerprint of the SAW--SW coupling mechanism at finite $|\vec{k}_{\mathrm{SAW}}|$. Moreover, along $\psi=180^\circ$ the interaction between SAW and backward SWs occurs at higher $f_{\textrm{SAW}}$ for the competing regime $B_0\sim B_u$ than for the Zeeman-dominated regime $B_0\gg B_u$ [see horizontal dashed line in Fig.~\ref{fig4:SAWSWParallelPumpingPsivsFsaw}]. This trend is consistent with a competition between Zeeman and uniaxial anisotropy energies. In the competing regime [Fig.~\ref{fig4:SAWSWParallelPumpingPsivsFsaw}(a)], the equilibrium magnetization is more easily reoriented and shifts the phase-matching condition to higher frequencies. In contrast, in the Zeeman-dominated regime [Fig.~\ref{fig4:SAWSWParallelPumpingPsivsFsaw}(b)] the magnetization follows $B_0$ more closely, so the dispersion (and hence the phase-matching conditions) varies more smoothly and the corresponding parallel SAW--SW region appears at lower $f_{\mathrm{SAW}}$.\\
\indent Taken together, Fig.~\ref{fig4:SAWSWParallelPumpingPsivsFsaw} identifies the optimal absorption resonance in the $\psi-f_{\mathrm{SAW}}$ plane for given $\{B_0,B_u,\varphi_u\}$ values. The regime $B_0\sim B_u$ yields the strongest and most structured anisotropy-controlled windows (i. e., sharp cusp-like onsets and split ridges), enabling efficient frequency-selective gating of magnetoacoustic pumping via field orientation. In contrast, the regime $B_0\gg B_u$ restores a more regular, dispersion dominated response with smoother bands, which is advantageous for stable operation and straightforward tuning. Consequently, $\Delta P(\psi,f_{\mathrm{SAW}})$ 2D maps serve as practical design charts for antenna-less magnonic pumping and reconfigurable acoustic spintronic functionality.
For a given $B_0$, the map therefore directly indicates which combinations of $\{\psi, f_{\textrm{SAW}}\}$ maximize absorption (or gain), and illustrates the expected $180^{\circ}$ periodicity in $\psi$ that can be used as an experimental fingerprint of the SAW--SW coupling mechanism.\\
\indent In conclusion, we performed a micromagnetic analysis of SAW--SW coupling at finite wave vector ($k_{\mathrm{SAW}}\neq 0$) using GPU-accelerated MuMax3 software, implementing a realistic magnetoacoustic excitation field $B_{\textrm{exc}}$ that includes both strain $\varepsilon_{\mu\nu}$ and lattice rotation $\omega_{\mu\nu}$ with a $\pi/2$ phase shift. By computing the absorbed magnetic power $\Delta P$ in the frequency domain, we identified robust symmetry fingerprints and ``branch-cut'' structures that emerge from the interplay between weak in-plane uniaxial anisotropy, magnetoacoustic drive and dipolar terms. We first studied the SAW--SW coupling dynamics, in the absence of uniaxial anisotropy ($B_u=0$). The coupling remained largely smooth and exhibited the expected $180^\circ$ periodicity with respect to the field orientation $\psi$ if we neglected non-reciprocity. While the inclusion of lattice rotation substantially enhanced the non-reciprocity and sharpened resonance sectors, it also allowed for the parallel SAW-SW interaction to emerge. 
Then, when a weak uniaxial anisotropy was introduced ($B_u\neq 0$), sharp and ``cusp''-like features appeared in SAW--SW coupling strength, reflecting sharp reorientation of the equilibrium magnetization under competition between Zeeman and anisotropy energies. The inclusion of $\omega_{\mu\nu}$ not only rescales the response; it selectively strengthens and restructures the anisotropy-induced resonance sectors. 
\indent From a practical standpoint, our results indicate that a weak in-plane uniaxial anisotropy and its orientation provide an efficient knob without changing the excitation frequency, with the strongest leverage obtained when $B_0$ is comparable to $B_u$. Increasing the SAW frequency enlarges the regions where parallel SAW--SW resonance is expected, and the resulting $\Delta P$ maps offer a direct blueprint for future experimental and device design. The interplay of magnetic anisotropy with strain, lattice rotation and dipolar energy gives rise to rich behavior. While the present study is numerical, the predicted symmetry constraints, nonreciprocity, $180^\circ$ periodicity in $\psi$ and systematic shifts of the $\Delta P$ contours with $\varphi_u$ and $f_{\mathrm{SAW}}$ provide concrete, testable scenarios for future SAW-driven SW experiments in thin magnetic films.
\section*{Acknowledgments}
The authors acknowledge the French National Research Agency (ANR) under contract N$^{o}$ ANR-22-CE24-0015 (SACOUMAD). The authors thank C. Gourdon and L.  Thevenard for fruitful discussions.
\bibliography{MagPhonSimulBib}
\appendix
\section{Magnetic power absorption used in MuMax3 simulations}
The SAW attenuation into the magnetic film is quantified by the magnetic power absorption per unit area [W/m$^2$]. In linear response, for a harmonic excitation at frequency $f=f_{\mathrm{SAW}}$, the magnetic power absorption is,
\begin{align}
\Delta P(f)
= -\pi f\ell_z\Im\!\left\{\tilde{\vec{B}}_{\mathrm{exc}}^{\dagger}(f)\cdot\overleftrightarrow{\chi}(f)\cdot\tilde{\vec{B}}_{\mathrm{exc}}(f)\right\},
\label{magnetipowerabsorption}
\end{align}
where tilded quantities denote the complex Fourier amplitudes in the frequency domain, $\chi$ is the Polder (frequency-dependent) susceptibility and $(.)^{\dagger}$ denotes the conjugate transpose operation. The susceptibility relates the reduced magnetization $\delta\tilde{\vec m}=\delta\tilde{\vec M}/M_s$ to the excitation field as,
\begin{align}
\delta\tilde{\vec m}(f)=\overleftrightarrow{\chi}(f)\cdot \tilde{\vec B}_{\mathrm{exc}}(f).
\label{chi_def}
\end{align}
Substituting Eq.~\eqref{chi_def} into Eq.~\eqref{magnetipowerabsorption} yields,
\begin{align}
\Delta P(f)
= -\pi fM_s\ell_z\Im\!\left\{\sum_{\alpha}\tilde{B}_{\mathrm{exc},\alpha}^{\dagger}(f)\,\delta\tilde{m}_{\alpha}(f)\right\},
\label{magnetipowerabsorption_compact}
\end{align}
with $\alpha\in\{x,y,z\}$. We write the complex Fourier amplitudes as (row vectors),
\begin{align}
\delta\tilde{m}_{\alpha}&=\delta\tilde{m}_{\alpha,r}+\mathrm{i}\delta\tilde{m}_{\alpha,i}, \nonumber \\
\tilde{B}_{\mathrm{exc},\alpha}&=\tilde{B}_{\mathrm{exc},\alpha,r}+\mathrm{i}\tilde{B}_{\mathrm{exc},\alpha,i},
\end{align}
where the subscripts $r\;\textrm{and}\;i$ stand for real and imaginary parts of the quantities. Defining $\tilde{B}_{\mathrm{exc},\alpha}^{\dagger}=\tilde{B}_{\mathrm{exc},\alpha,r}-\mathrm{i}\tilde{B}_{\mathrm{exc},\alpha,i}$ (a column vector), we obtain, 
\begin{align}
\tilde{B}_{\mathrm{exc},\alpha}^{\dagger}\,\delta\tilde{m}_{\alpha}
&=(\tilde{B}_{\mathrm{exc},\alpha,r}-\mathrm{i}\tilde{B}_{\mathrm{exc},\alpha,i})
(\delta\tilde{m}_{\alpha,r}+\mathrm{i}\delta\tilde{m}_{\alpha,i}) \nonumber\\
&=\tilde{B}_{\mathrm{exc},\alpha,r}\delta\tilde{m}_{\alpha,r}
+\tilde{B}_{\mathrm{exc},\alpha,i}\delta\tilde{m}_{\alpha,i} \nonumber\\
&+\mathrm{i}\left(\tilde{B}_{\mathrm{exc},\alpha,r}\delta\tilde{m}_{\alpha,i}
-\tilde{B}_{\mathrm{exc},\alpha,i}\delta\tilde{m}_{\alpha,r}\right).
\end{align}
Therefore,
\begin{align}
\Im\!\left\{\tilde{B}_{\mathrm{exc},\alpha}^{\dagger}\,\delta\tilde{m}_{\alpha}\right\}
=\tilde{B}_{\mathrm{exc},\alpha,r}\delta\tilde{m}_{\alpha,i}
-\tilde{B}_{\mathrm{exc},\alpha,i}\delta\tilde{m}_{\alpha,r},
\end{align}
and Eq.~\eqref{magnetipowerabsorption_compact} transforms into the component-wise equation,
\begin{equation}
\resizebox{0.9\linewidth}{!}{$
\begin{aligned}
        \Delta P(f) = - \pi f M_s \ell_z 
        \sum_{\alpha\in {x,y,z}} \left( 
        \tilde{B}_{\textrm{exc},\alpha,r}\delta \tilde{m}_{\alpha,i} - \tilde{B}_{\textrm{exc},\alpha,i}\delta \tilde{m}_{\alpha,r} 
        \right).
\end{aligned}$}
\label{magnetipowerabsorptionfinaldetailledappendix}
\end{equation}
We calculate the complex amplitudes in the stationary regime at $t=N/f_{\mathrm{SAW}}$ (where $N$ is a sufficiently large integer to reach the stationary regime) over a full SAW wavelength. We write,
\begin{align}
\delta m_\alpha(x,t)&=\Re\!\left[\delta\tilde m_\alpha\,e^{\mathrm{i}(kx-2\pi f t)}\right]\Big|_{t=N/f_{\textrm{SAW}}},\nonumber\\
B_{\mathrm{exc},\alpha}(x,t)&=\Re\!\left[\tilde B_{\mathrm{exc},\alpha}\,e^{\mathrm{i}(kx-2\pi f t)}\right]\Big|_{t=N/f_{\textrm{SAW}}},
\end{align}
where the complex amplitudes are obtained (at a fixed reference phase at $f_{\textrm{SAW}}$) from spatial projections over one acoustic wavelength $\lambda_{\textrm{SAW}}=2\pi/k_{\textrm{SAW}}$:
\begin{equation}
\begin{cases}
\delta \tilde{m}_{\alpha,r} = \dfrac{2}{\lambda_{\mathrm{SAW}}}\displaystyle\int_{0}^{\lambda_{\mathrm{SAW}}}\delta m_\alpha(x)\cos(kx)\,dx,\\[6pt]
\delta \tilde{m}_{\alpha,i} = \dfrac{2}{\lambda_{\mathrm{SAW}}}\displaystyle\int_{0}^{\lambda_{\mathrm{SAW}}}\delta m_\alpha(x)\sin(kx)\,dx,\\[10pt]
\tilde{B}_{\mathrm{exc},\alpha,r} = \dfrac{2}{\lambda_{\mathrm{SAW}}}\displaystyle\int_{0}^{\lambda_{\mathrm{SAW}}}B_{\mathrm{exc},\alpha}(x)\cos(kx)\,dx,\\[6pt]
\tilde{B}_{\mathrm{exc},\alpha,i} = \dfrac{2}{\lambda_{\mathrm{SAW}}}\displaystyle\int_{0}^{\lambda_{\mathrm{SAW}}}B_{\mathrm{exc},\alpha}(x)\sin(kx)\,dx.
\end{cases}
\label{equationsSAWSW}
\end{equation}
\end{document}